# The Analogy between Electromagnetics and Hydrodynamics

Kai-Xin Hu[a,1]

[a]Zhejiang Provincial Engineering Research Center for the Safety of Pressure Vessel and Pipeline, Key Laboratory of Impact and Safety Engineering, Ministry of Education, School of Mechanical Engineering and Mechanics, Ningbo University, Ningbo, Zhejiang 315211, China

## Abstract

The similarity between electromagnetics and hydrodynamics has been noticed for a long time. Maxwell developed an analogy, where the magnetic field and the vector potential in electromagnetics are compared to the vorticity and velocity in hydrodynamics, respectively. However, this theory cannot make a correspondence of energy between two subjects. In the present work, the electromagnetic fields in a conducting medium are compared to the flow fields of an incompressible Newtonian fluid. The result shows that the magnetic induction intensity, current density, Lorenz force, superconductor boundary, Ohm's law and Ampere force in electromagnetics are analogous to the velocity, vorticity, Lamb vector, solid boundary, Newton's law of viscosity and Kutta-Joukowski theorem of lift force in hydrodynamics, respectively. The Navier-Stokes equation is derived for the evolution of magnetic field in the medium by using the Maxwell equations, Lorenz force and Ohm's law. The work is useful for a deep understanding of electromagnetics and hydrodynamics.

[1]Email:hukaixin@nbu.edu.cn

## 1. Introduction

The method of physical analogy, which was first proposed by Maxwell, exploits the formal similarities between two distinct domains of science[1]. It can bring a better understanding of different fields and has greatly promoted the development of physics in history. For example, W.Thomson [2] has solved the problem in electrostatics by making analogy between the electrostatic force and the heat flux, while Maxwell [3] has developed the equations of electromagnetics on the vortex-idle wheel model of electromagnetic medium. Feynmann [4] has also shown many examples of analogies between different physical problems in his lectures.

The similarity between electromagnetics and hydrodynamics has been noticed since the 19[th] century. Helmholtz has taken the rotating fluid element to be analogous to the electric current, where the velocity field **v** is analogous to the magnetic induction intensity **B** [1]. He has used the Biot-Savart formula of electromagnetism to derive the velocity from the vorticity [5]. However, Maxwell [3] believes that Helmholtz has misconstrued the corresponding analogy, and has proposed another analogy, where the magnetic field and the electric current in electromagnetics are compared to the rotation and motion of fluids in hydrodynamics, respectively. J.J.Thomson [6] has made analogy between the lines of electric force in vacuum and vortex filaments in an inviscid fluid.

In recent years, the analogy proposed by Maxwell has been developed by many authors. Marmanis [7] has initiated a new theory of turbulence based on the analogy between electromagnetism and turbulent hydrodynamics. The magnetic and electric fields $\mathbf{B}, \mathbf{E}$ correspond to the vorticity $\boldsymbol{\omega} = \nabla \times \mathbf{v}$ and Lamb vector $\mathbf{L} = \boldsymbol{\omega} \times \mathbf{v}$ in an inviscid fluid flow, respectively. Later, this analogy is adopted by Martins & Pinheiro [8], Büker & Tripoli [9] and Panakkal, Parameswaran & Vedan [10]. Scofield and Huq [11] have presented the fluid dynamical analogs of the Lorentz force law and Poynting theorem of electrodynamics, where the vorticity in geometrodynamical theory of fluids is analogous to the magnetic field.

A new analogy between the equations of compressible fluids and Maxwell equations has been presented by Kambe [12,13]. The magnetic field is compared to the vorticity, the electric field intensity $\mathbf{E}$ to the convective acceleration $\mathbf{a}_c = (\mathbf{v} \cdot \nabla)\mathbf{v}$, and the sound wave is analogous to the electromagnetic wave. This work was followed by Jamati [14], who has made analogy between vortex waves and electromagnetic waves. Arbab [15] has made analogy between electromagnetics and hydrodynamics, where the magnetic field still plays a role of vorticity while the electric field corresponds to the hydroelectric field $\mathbf{E}_h = -\dfrac{\partial \mathbf{v}}{\partial t} - \nabla\left(\dfrac{1}{2}|\mathbf{v}|^2\right)$.

However, there are some disadvantages in these analogies. The most important one is the energy. In electromagnetics, the energy density of electromagnetic field is proportional to $\dfrac{1}{2}\left(|\mathbf{B}|^2 + \dfrac{1}{c^2}|\mathbf{E}|^2\right)$, where $c$ is the speed of light. On the contrary, in hydrodynamics, $\dfrac{1}{2}|\boldsymbol{\omega}|^2$ does not correspond to a kind of energy density in the flow. Indeed, it is called the enstrophy density in the vortex dynamics [16], and its volume integral measures the dissipation rate of kinetic energy. Similarly, none of $\dfrac{1}{2}|\mathbf{L}|^2, \dfrac{1}{2}|\mathbf{a}_c|^2, \dfrac{1}{2}|\mathbf{E}_h|^2$ corresponds to a kind of energy density. Therefore, we cannot make a proper correspondence of energy in these analogies.

Another serious problem is that these analogies impose some special relationships between the magnetic and electric fields. The analogy $\mathbf{B} \sim \boldsymbol{\omega}$, $\mathbf{E} \sim \mathbf{L} = \boldsymbol{\omega} \times \mathbf{v}$ leads to $\mathbf{B} \perp \mathbf{E}$. When $\mathbf{E} \sim \mathbf{a}_c$ or $\mathbf{E} \sim \mathbf{E}_h$, the magnetic and electric fields are determined by a three-dimensional vector $\mathbf{v}$, which differs from the vector potential $\mathbf{A}$ in the classical theory. However, these relations are not generally found in electromagnetism. Although the analogies above have achieved some progress and made applications in acoustics [12] and turbulence [8], these obvious problems restrict their further development and application.

In the present work, the analogy between electromagnetics and hydrodynamics is developed by following the way of Helmholtz, where the velocity $\mathbf{v}$ and the vorticity $\boldsymbol{\omega}$ are analogous to the magnetic field $\mathbf{B}$ and the current density $\mathbf{j}$, respectively. We show that this analogy can construct a variety of correspondences between physical quantities.

The paper is organized as follows. In Section 2, the similarities of equations and field patterns between electromagnetic and flow fields are presented. The Ohm's law of resistance is analogous to the Newton's Law of viscosity, and the equation of Ampere force is analogous to Kutta-Joukowski theorem of lift. The electromagnetic field in a conducting medium is analyzed. Then in Section 3, the Navier-Stokes equation is derived from the Maxwell equations, Lorenz force and Ohm's law. After that, some topics are discussed in Section 4. Finally, the conclusions are drawn in Section 5.

## 2. The similarities between electromagnetics and hydrodynamics

In this section, we list some similarities between the electromagnetic field and the incompressible fluid flow in the following paragraphs.

### 2.1 The governing equations of magnetic field and flow field

In electromagnetics, the magnetic field is divergence-free,

$$\nabla \cdot \mathbf{B} = 0. \tag{2.1}$$

The Ampere's law of magnetostatics is

$$\oint_C \mathbf{B} \cdot \mathrm{d}\mathbf{l} = \mu I, \tag{2.2}$$

where $I$ is the total current through the loop $C$ and $\mu$ is the permeability. Its differential form is

$$\nabla \times \mathbf{B} = \mu \mathbf{j}, \tag{2.3}$$

where $\mathbf{j}$ is the current density.

On the other hand, in hydrodynamics, the continuity equation of an incompressible fluid is

$$\nabla \cdot \mathbf{v} = 0. \tag{2.4}$$

The circulation $\Gamma$ around a loop $C'$ is defined as

$$\oint_{C'} \mathbf{v} \cdot \mathrm{d}\mathbf{l} = \Gamma, \tag{2.5}$$

while the vorticity $\boldsymbol{\omega}$ has

$$\nabla \times \mathbf{v} = \boldsymbol{\omega}. \tag{2.6}$$

Based on (2.1)-(2.6), we observe that $\mathbf{B}, I, \mathbf{j}$ in electromagnetics are analogous to $\mathbf{v}, \Gamma, \boldsymbol{\omega}$ in hydrodynamics.

### 2.2 The patterns of magnetic field and flow field

The patterns of magnetic and flow fields are similar in many cases. The magnetic field induced by a solenoid is displayed in Fig.1(a). When the number of turns in the wind per length along the cylinder is large enough, the magnetic field lines are restricted inside the cylinder except the region near two ends, which are similar to the streamlines in a pipe flow in Fig.1(b).

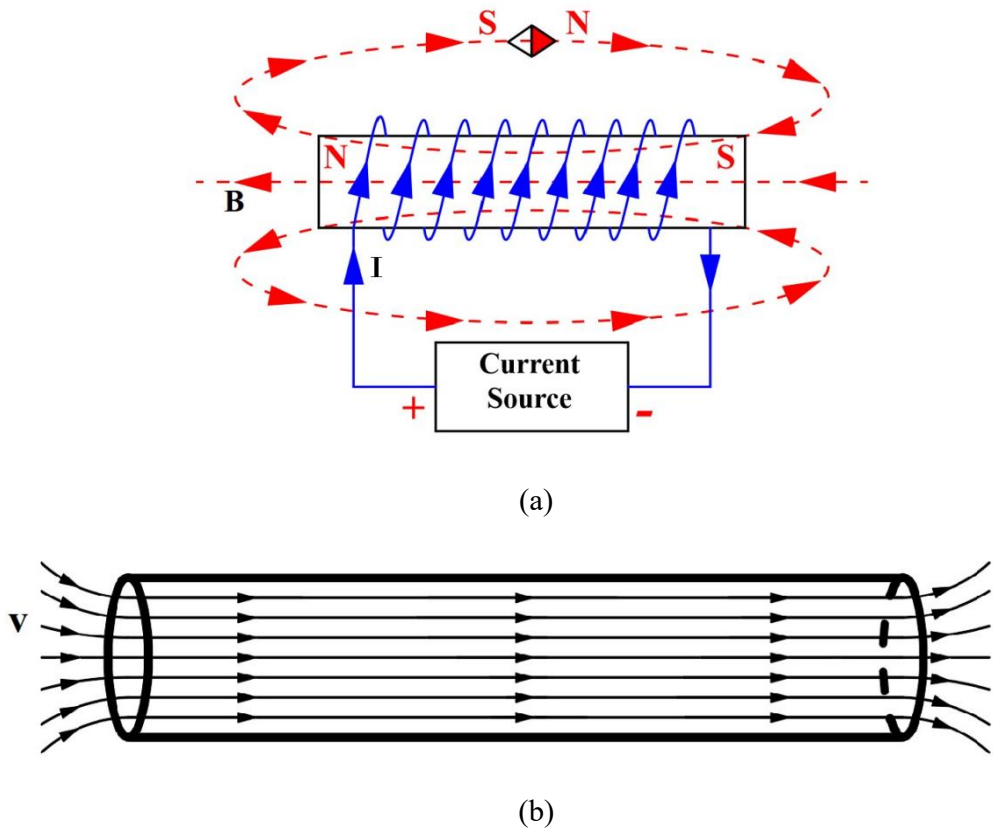

(a)

(b)

Fig. 1.   (a)The magnetic field induced by a solenoid. (b)The fluid flow in a pipe.

In Fig.2(a), the magnetic field lines induced by a current filament form concentric circles. The magnetic field has $\mathbf{B} = \dfrac{\mu I}{2\pi r} \mathbf{e}_\theta$ , where $\mathbf{e}_\theta$ is the unit vector in the circumferential direction. The corresponding one in fluid flow is the point vortex in Fig.2(b), whose velocity field has $\mathbf{v} = \dfrac{\Gamma}{2\pi r} \mathbf{e}_\theta$ . Similarly, the magnetic field induced by a current ring in Fig.3(a) is like the velocity field induced by a vortex ring in Fig.3(b).

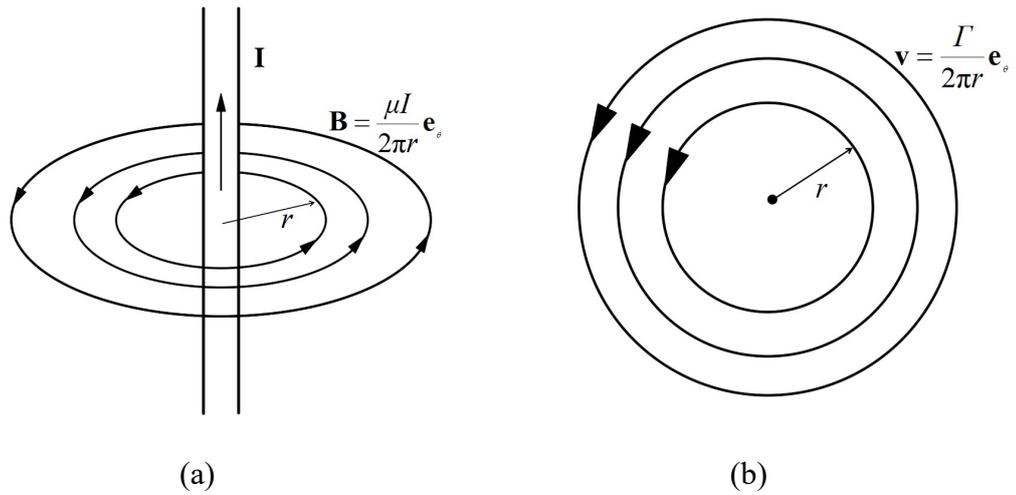

(a)                                                    (b)

Fig. 2. (a) The magnetic field around a current filament; (b) The velocity field of a point

vortex.

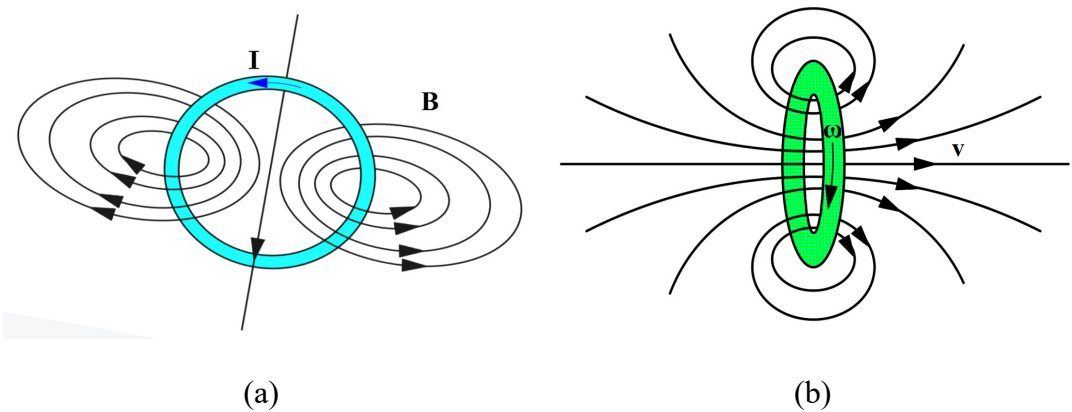

(a)                                                    (b)

Fig. 3. (a) The magnetic field induced by a current ring; (b) The velocity field induced by a vortex ring.

Fig.4(a) shows the Meissner effect, which is the expulsion of magnetic field from a conductor when it is cooled below the critical temperature $T_c$ and transits to the superconducting state. The magnetic field lines around the superconductor are similar to the streamlines around a sphere in the fluid flow, where the streamlines cannot penetrate the solid boundary. This similarity has been used by Faber[17], where he presented a magnetostatic analogue for vorticity-free flows to solve flow problems.

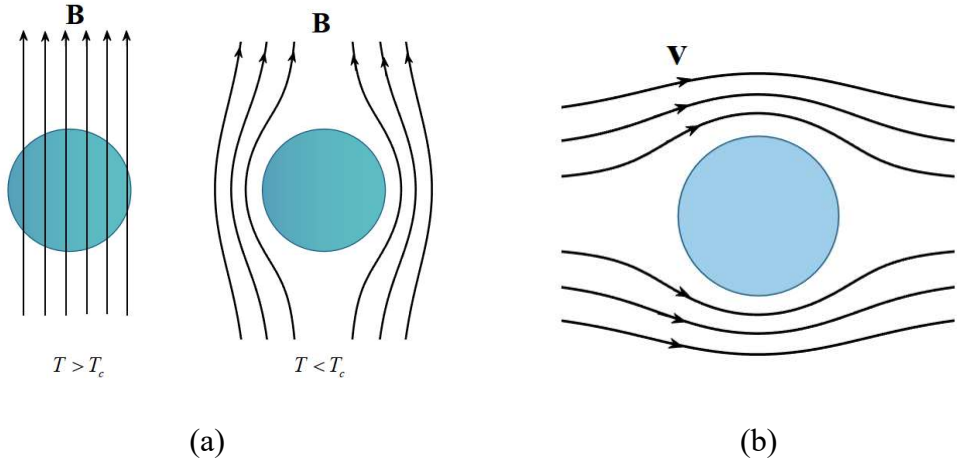

Fig. 4. (a) The magnetic field around a conductor; (b) The velocity field around a sphere.

## 2.3   Ohm's law of resistance and Newton's Law of Viscosity

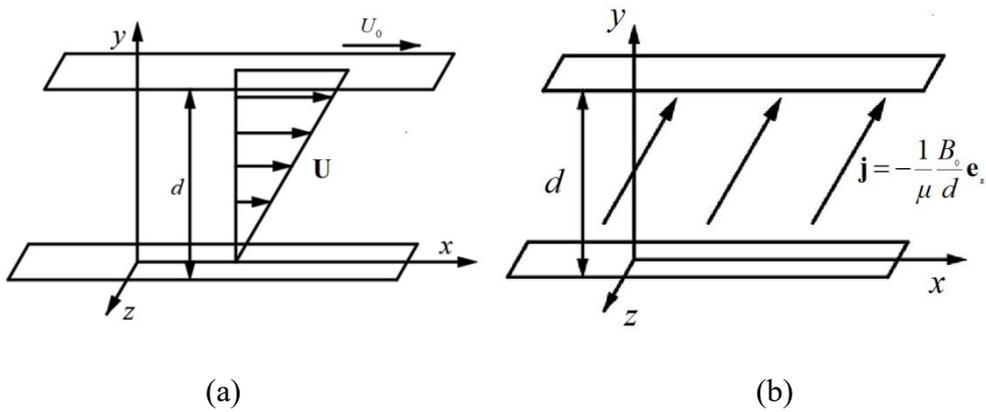

(a)                                    (b)

Fig. 5.   (a)The plane Couette flow; (b) An uniform current field between two

parallel plates.

We consider the plane Couette flow in Fig.5(a), where the fluid between two parallel plates is driven by the upper plate at the speed of $U_0$ while the lower plate is static. Newton's law of viscosity suggests that

$$\tau_{12} = \eta U_0 / d, \tag{2.7}$$

where $\tau_{12}$ is the shear stress, $\eta$ is the dynamic viscosity and $d$ is the distance between two plates. Thus, the viscous dissipation rate per unit volume is

$$\phi_{\mathrm{f}} = \eta\left(U_0 / d\right)^2. \tag{2.8}$$

We compare the velocity field $\mathbf{U} = \dfrac{U_0 y}{d}\mathbf{e}_x$ of plane Couette flow to a magnetic field

$\mathbf{B} = \dfrac{B_0 y}{d}\mathbf{e}_x$, where its current density is $\mathbf{j} = -\dfrac{1}{\mu}\dfrac{B_0}{d}\mathbf{e}_z$ (see in Fig.5(b)). Suppose the

current is driven by an electric field $\mathbf{E}$, Ohm's law of resistance suggests that

$$\mathbf{j} = \sigma\mathbf{E}, \tag{2.9}$$

where $\sigma$ is the electrical conductivity. The dissipation rate per unit volume is

$$\phi_{\mathrm{e}} = \mathbf{j} \cdot \mathbf{E} = \frac{1}{\sigma}\mathbf{j}^2 = \frac{1}{\sigma\mu^2}\left(\frac{B_0}{d}\right)^2. \tag{2.10}$$

Therefore, the fluid viscosity $\eta$ is analogous to the electrical resistivity $\chi = 1/\sigma$. It should be noted that the current can also be driven by other non-electrostatic forces,

such as in the cases of battery and plasmas. So the general form of Ohm's law can be

written as

$$\mathbf{j} = \sigma\left(\mathbf{E} + \mathbf{K}\right), \tag{2.11}$$

where $\mathbf{K}$ stands for another force field. This will be used in Section 3.

## 2.4   Ampere force and Kutta-Joukowski theorem

The Lorenz force on a moving charge $q$ with velocity $\mathbf{v}$ in the magnetic field is

$$\mathbf{f}_q = q\mathbf{v} \times \mathbf{B}. \tag{2.12}$$

So the Lorenz force density is $\mathbf{f} = \mathbf{j} \times \mathbf{B}$, which is analogous to the Lamb vector

$\mathbf{L} = \boldsymbol{\omega} \times \mathbf{v}$ in hydrodynamics.

For a current filament with length $l$ in the magnetic field, the Lorenz force on

moving charges lead to the Ampere force on the filament

$$\mathbf{F} = \mathbf{I} \times \mathbf{B}l. \tag{2.13}$$

We consider a current filament imposed in an external magnetic field in Fig.5,

where $\mathbf{B}_1$ is the magnetic induction intensity of external field in the place of the filament, and **i** is the current of filament.

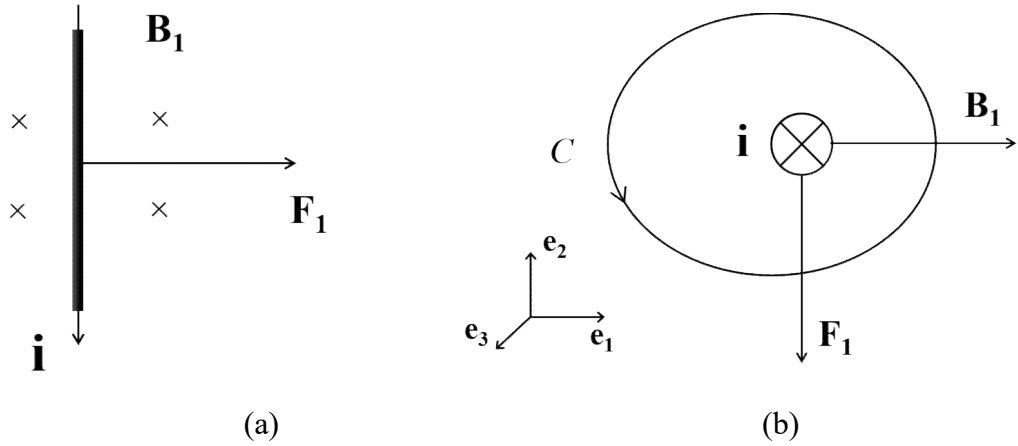

(a)                                                                 (b)

Fig. 5   The Ampere force on a current filament：(a)side view；(b) top view.

Due to the Ampere's law, the loop-integration of magnetic field around the filament is

$$\oint_C \mathbf{B} \cdot \mathbf{dl} = \mu |\mathbf{i}|,$$

(2.14)

where $C$ is loop. The Ampere force on the filament per unit length is

(2.15)

$$\mathbf{F}_1 = -\mathbf{B}_1 \times \mathbf{i},$$

where $\mathbf{i} = -|\mathbf{i}|\mathbf{e}_3$ .

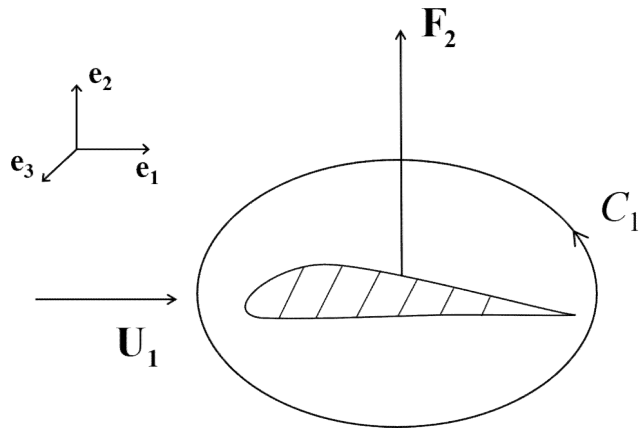

Fig. 6    The two-dimensional incompressible irrotational flow around an aerofoil.

In hydrodynamics, the Kutta-Joukowski theorem shows that the lift force on an

aerofoil in the two-dimensional incompressible irrotational flow is

$$\mathbf{F}_2 = \rho \mathbf{U}_1 \times \mathbf{\Gamma}, \tag{2.16}$$

where $\mathbf{F}_2$ is the lift force per unit length, $\rho$ is the fluid density, $\mathbf{U}_1$ is the velocity of the fluid at infinity, while the aerofoil is static. $\mathbf{\Gamma}$ is the circulation of the loop $C_1$ around the airfoil,

$$\mathbf{\Gamma} = \left( \oint_{C_1} \mathbf{v} \cdot \mathbf{dl} \right) \mathbf{e}_3. \tag{2.17}$$

It can be seen that (2.15) is similar to (2.16), while the direction of $\mathbf{F}_1$ is just opposite to $\mathbf{F}_2$. So we can make an analogy between $\left( -\mathbf{F}_1 \right)$ and $\mathbf{F}_2$, where $\mathbf{B}$ and $\mathbf{i}$ are analogous to $\mathbf{v}$ and $\mathbf{\Gamma}$, respectively. A similar analogue has been proposed by Faber[17], where the Ampere force on a current-carrying wire is compared to the Magnus force on a spinning cylinder.

## 2.5 The electromagnetic field in a conducting medium

The Maxwell equations, which are the governing equations of electromagnetic phenomena, are displayed as follows,

$$\nabla \cdot \mathbf{B} = 0, \tag{2.18a}$$

$$\nabla \times \mathbf{E} = -\frac{\partial}{\partial t} \mathbf{B}, \tag{2.18b}$$

$$\nabla \cdot \mathbf{D} = \vartheta, \tag{2.18c}$$

$$\nabla \times \mathbf{H} = \mathbf{j} + \frac{\partial}{\partial t} \mathbf{D}, \tag{2.18d}$$

where $\vartheta$ is the charge density, $\mathbf{D}$ and $\mathbf{H}$ are the electric displacement and magnetic field intensity, respectively.

We consider the electromagnetic fields in a conducting medium. Suppose that $\mathbf{D} = \varepsilon \mathbf{E}$ and $\mathbf{H} = \mathbf{B} / \mu$, where $\varepsilon$ is the permittivity. The relaxation time for the charge in the medium is $\tau = \varepsilon / \sigma$. When the frequency scale of electromagnetic fields $\omega$ is much smaller than $1/\tau$, that is

$$\frac{\sigma}{\varepsilon\omega} >> 1, \tag{2.19}$$

then, there is no bulk charge density inside the medium. (2.18c) can be simplified to

$$\nabla \cdot \mathbf{E} = 0. \tag{2.20}$$

In addition, we suppose that

$$\frac{c}{l\omega} >> 1, \tag{2.21}$$

where $c = 1/\sqrt{\mu\varepsilon}$ is the speed of light in the medium and $l$ is the length scale. In this condition, we compare the magnitude of electric field and magnetic induction by the dimensionless ratio,

$$\frac{|\mathbf{E}|}{c|\mathbf{B}|} = O\left(\frac{\omega l}{c}\right) << 1, \tag{2.22}$$

which means that the fields are predominantly magnetic [18]. The energy density of electromagnetic fields in a medium is $e_e = \frac{1}{2}\left(\varepsilon|\mathbf{E}|^2 + \frac{1}{\mu}|\mathbf{B}|^2\right)$. (2.22) suggests that the energy of electric field can be neglected, so

$$e_e = \frac{1}{2\mu}|\mathbf{B}|^2. \tag{2.23}$$

Similarly, the Poynting vector $\mathbf{S} = \frac{1}{\mu}\mathbf{E} \times \mathbf{B}$, which shows the energy flux density, can also be neglect in the equation of stress. Meanwhile, (2.21) means that the fields are quasi-static, so (2.18d) can be simplified to

$$\nabla \times \mathbf{B} = \mu\dot{\mathbf{j}}. \tag{2.24}$$

Moffatt[19] has established a mathematical analogy between the Euler equations for steady flow of an inviscid incompressible fluid

$$0 = \nabla h + \boldsymbol{\omega} \times \mathbf{v}, \quad \boldsymbol{\omega} = \nabla \times \mathbf{v}, \quad \nabla \cdot \mathbf{v} = 0, \qquad (2.25)$$

and the equations of magnetostatic equilibrium in a perfectly conducting fluid

$$0 = -\nabla s + \mathbf{j}' \times \mathbf{B}, \quad \mathbf{j}' = \nabla \times \mathbf{B}, \quad \nabla \cdot \mathbf{B} = 0, \qquad (2.26)$$

where $h$ is the Bernoulli function in the inviscid fluid and $s$ is the pressure field in the conducting fluid. The analogy is evidently between the variables

$$\mathbf{v} \leftrightarrow \mathbf{B}, \quad \boldsymbol{\omega} \leftrightarrow \mathbf{j}', \quad h \leftrightarrow s_0 - s, \qquad (2.27)$$

where $s_0$ is a constant.

This analogy can be generalized to the unsteady case in the medium with energy dissipation. In Section 3, we derive the governing equation for the quasi-static electromagnetic fields in a conducting medium, which is similar to the Navier-Stokes equation.

In order to make the analogy more clearly, we would like to change the dimension of $\mathbf{B}$ to that of the velocity. This can be done by dividing a constant $\Theta$ to all variables $\mathbf{B}, \mathbf{E}, \mathbf{j}$ in Maxwell equations. $\Theta$ can be determined as follows.

The energy density in a fluid flow is

$$e_f = \frac{1}{2} \rho |\mathbf{v}|^2. \qquad (2.28)$$

Let $\mathbf{B} = \Theta \mathbf{v}$. If the energy density in (2.23) equals that in (2.28), we have

$$\Theta^2 / \mu = \rho. \qquad (2.29)$$

On the other hand, let $B_0 = \Theta U_0$ in (2.10), the dissipation in (2.8) equals to that in (2.10), which requires that

$$\eta = \frac{1}{\sigma \mu^2} \Theta^2. \qquad (2.30)$$

Thus,

$$\nu = \frac{\eta}{\rho} = \frac{1}{\sigma \mu}, \tag{2.31}$$

where $\nu$ is the kinematic viscosity. In the following, all variables of electromagnetic fields have been divided $\Theta$, so $\mathbf{B}$ has the dimension of velocity.

## 3. The derivation of Navier-Stokes equation

Now we consider the electromagnetic fields in a conducting medium. The curl of (2.24) shows that

$$\nabla^2 \mathbf{B} = -\mu \nabla \times \mathbf{j}. \tag{3.1}$$

Suppose the electric current is driven by the electric field $\mathbf{E}$ and a non-electrostatic force field $\mathbf{K}$, then we substitute the Ohm's law (2.11) in to (3.1) and derive

$$\nabla^2 \mathbf{B} = -\mu \sigma \nabla \times (\mathbf{E} + \mathbf{K}). \tag{3.2}$$

Then, substituting (2.18b) and (2.31) into (3.2), the following equation can be obtained,

$$\nu \nabla^2 \mathbf{B} = \frac{\partial}{\partial t} \mathbf{B} - \nabla \times \mathbf{K}. \tag{3.3}$$

We do not discuss the micro-mechanism of $\mathbf{K}$, but pay attention to its macro-effect. The dimension of $(-\nabla \times \mathbf{K})$ is the same as that of $\frac{\partial}{\partial t} \mathbf{B}$, which suggests that this term refers to a kind of force per unit mass on the control volume.

In electromagnetics, the force on a control volume comes from three parts. The first one is the force on electric charges in the volume. As there is no bulk charge density inside the medium, the Lorenz force is

$$\mathbf{f}' = \mu \mathbf{j} \times \mathbf{B} = (\nabla \times \mathbf{B}) \times \mathbf{B}, \tag{3.4}$$

which makes a contribution to $(-\nabla \times \mathbf{K})$.

The second one is caused by the variation of energy flux $\dfrac{\partial}{\partial t}\mathbf{S}$. The magnitude of

$\dfrac{\partial}{\partial t}\mathbf{S}$ and $\mathbf{f}'$ is compared by the dimensionless ratio $\left|\dfrac{\partial}{\partial t}\mathbf{S}\right|/\left|\mathbf{f}'\right| = O\left(\left|\dfrac{\omega l}{c}\right|^2\right) << 1$. So

this one can be neglected.

The third part comes from the medium outside the volume, which can be expressed as the gradient of a scalar $\nabla p'$. Here, the stress in the medium is supposed to be isotropic. Therefore, we have

$$-\nabla \times \mathbf{K} = (\nabla \times \mathbf{B}) \times \mathbf{B} + \nabla p'. \qquad (3.5)$$

From the mathematical point of view, (3.5) gives the Helmholtz decomposition of $(\nabla \times \mathbf{B}) \times \mathbf{B}$ (Lorenz force density), where $(-\nabla p')$ is the curl-free vector field and $(-\nabla \times \mathbf{K})$ is the divergence-free vector field. Physically, (3.5) shows that the Lorentz force is responsible for motional electromotive force.

Substituting (3.5) into (3.3), we have

$$\nu \nabla^2 \mathbf{B} = \frac{\partial}{\partial t}\mathbf{B} + (\nabla \times \mathbf{B}) \times \mathbf{B} + \nabla p'. \qquad (3.6)$$

By using the divergence of $\mathbf{B}$ in (2.1), the following identity is obtained,

$$(\nabla \times \mathbf{B}) \times \mathbf{B} + \frac{1}{2}|\mathbf{B}|^2 = (\mathbf{B} \cdot \nabla)\mathbf{B}. \qquad (3.7)$$

Then, the Navier-Stokes equation is derived as follows

$$\frac{\partial}{\partial t}\mathbf{B} + (\mathbf{B} \cdot \nabla)\mathbf{B} = -\nabla p + \nu \nabla^2 \mathbf{B}, \qquad (3.8)$$

where $p = p' - \dfrac{1}{2}|\mathbf{B}|^2$. It can be seen that $p' = p + \dfrac{1}{2}|\mathbf{B}|^2$ corresponds to the total pressure in hydrodynamics, while (3.6) is the Lamb form of Navier-Stokes equation.

## 4. Discussion

We discuss the some topics of the analogy in this section.

## 4.1 The viscous dissipation and the enstrophy

In hydrodynamics, the viscous dissipation of a Newtonian fluid per unit volume is

$$\phi = \frac{1}{2}\eta S_{ij}S_{ij}, \tag{4.1}$$

where $S_{ij} = \dfrac{\partial v_i}{\partial x_j} + \dfrac{\partial v_j}{\partial x_i}$. It can be obtained that

$$\frac{1}{4}S_{ij}S_{ij} = \frac{1}{4}\left(\frac{\partial v_i}{\partial x_j} - \frac{\partial v_j}{\partial x_i}\right)^2 + \frac{\partial v_i}{\partial x_j}\frac{\partial v_j}{\partial x_i} = \frac{1}{2}|\boldsymbol{\omega}|^2 + \nabla\cdot\left[(\mathbf{v}\cdot\nabla)\mathbf{v}\right]. \tag{4.2}$$

Therefore, the total dissipation is

$$\varPhi = \int \frac{1}{2}\eta S_{ij}S_{ij}\,\mathrm{dV} = 2\eta\left(\int \frac{1}{2}|\boldsymbol{\omega}|^2\,\mathrm{dV} + \int \mathbf{n}\cdot(\mathbf{v}\cdot\nabla)\mathbf{v}\mathrm{dS}\right), \tag{4.3}$$

where $\mathbf{n}$ is the outward unit normal vector on the boundary.

If the domain is the whole space and the velocity decays sufficiently rapidly at infinity, or $\mathbf{n}\cdot(\mathbf{v}\cdot\nabla)\mathbf{v} = 0$ on the boundary, then the second term on the right side of (4.3) vanishes [20]. This condition is satisfied in many laminar and turbulent flows, such as channel flows, pipe flows and jet flows. In these cases, the enstrophy $Q_{\mathrm{f}} = \int \frac{1}{2}|\boldsymbol{\omega}|^2\,\mathrm{dV}$ can measure the dissipation rate of kinetic energy, which is analogous to the dissipation rate caused by the electric resistance $Q_{\mathrm{e}} = \int \frac{1}{\sigma}\mathbf{j}^2\mathrm{dV}$. However, in some cases, the above analogy of dissipation is inappropriate. For example, a fluid making a solid-like rotation has no viscous dissipation, while $Q_{\mathrm{f}} > 0$.

## 4.2 The stress tensor

The stress tensor $\boldsymbol{\sigma}$ of a electromagnetic field is

$$\sigma_{ij} = \frac{1}{c^2}\left(E_iE_j - \frac{1}{2}\delta_{ij}|\mathbf{E}|^2\right) + \left(B_iB_j - \frac{1}{2}\delta_{ij}|\mathbf{B}|^2\right), \tag{4.4}$$

which is also called as the Maxwell stress tensor in electromagnetics. As the energy of electric field is negligible in the medium, $\boldsymbol{\sigma}$ can be simplified to

$$\sigma_{ij} = \left( B_i B_j - \frac{1}{2}\delta_{ij}|\mathbf{B}|^2 \right).\tag{4.5}$$

Then, it can be derived that

$$\nabla \cdot \boldsymbol{\sigma} = (\mathbf{B} \cdot \nabla)\mathbf{B} - \nabla\left(\frac{1}{2}|\mathbf{B}|^2\right) = (\nabla \times \mathbf{B}) \times \mathbf{B},\tag{4.6}$$

which is the Lorenz force in (3.4). In addition,

$$(\mathbf{B} \cdot \nabla)\mathbf{B} = \nabla \cdot \boldsymbol{\tau},\tag{4.7}$$

where $\tau_{ij} = B_i B_j$ corresponds to the stress tensor caused by the velocity field in a fluid flow. Thus, the difference between $\boldsymbol{\sigma}$ and $\boldsymbol{\tau}$ is $\frac{1}{2}\delta_{ij}|\mathbf{B}|^2$, which corresponds to the dynamic pressure in hydrodynamics.

### 4.3  The Current filament and Vortex filament

In hydrodynamics, the helicity $H_f$ is defined as[21]

$$H_f = \int \boldsymbol{\omega} \cdot \mathbf{v}\mathrm{dV},\tag{4.8}$$

which shows the topological structure of vortex filaments (twist and knot). Similarly, the current helicity $H_c$ and magnetic helicity $H_m$ are defined in electromagnetics as follows[22],

$$H_c = \int \mathbf{j} \cdot \mathbf{B}\mathrm{dV},\tag{4.9a}$$

$$H_m = \int \mathbf{B} \cdot \mathbf{A}\mathrm{dV}.\tag{4.9b}$$

Topologically, $H_c$ ( $H_m$ ) shows the linkage of current filaments (magnetic field lines). These two definitions can describe the magnetic activity on the solar surface and have been widely used in astrophysics[22]. In our analogy, the current filaments, $H_e$ and $\mathbf{B}$ correspond to the vortex filaments, $H_f$ and $\mathbf{v}$, respectively. On the other hand, if $H_m$ is compared to $H_f$, then $\mathbf{B}$ is analogous to $\boldsymbol{\omega}$, which keeps the same as the analogy proposed by Maxwell [3].

We consider the Ampere forces between parallel current filaments. The force per unit length on filament $\alpha$ is

$$\mathbf{f}_\alpha = \left(f_{\alpha x}, f_{\alpha y}\right) = \left(-I_\alpha B_{\alpha y}, I_\alpha B_{\alpha x}\right), \tag{4.10}$$

where the filament $\alpha$ at $\left(x_\alpha, y_\alpha\right)$ has the current $I_\alpha$, $\alpha = 1 \sim N$, $\mathbf{B}_\alpha = \left(B_{\alpha x}, B_{\alpha y}\right)$ is the magnetic induction intensity induced by other current filaments,

$$B_{\alpha x} = -\sum_{\beta \neq \alpha} \frac{\mu I_\beta}{2\pi r_{\alpha\beta}^2}\left(y_\alpha - y_\beta\right), \quad B_{\alpha y} = \sum_{\beta \neq \alpha} \frac{\mu I_\beta}{2\pi r_{\alpha\beta}^2}\left(x_\alpha - x_\beta\right), \tag{4.11}$$

and $r_{\alpha\beta} = \sqrt{\left(x_\alpha - x_\beta\right)^2 + \left(y_\alpha - y_\beta\right)^2}$. Therefore,

$$f_{\alpha x} = -I_\alpha \sum_{\beta \neq \alpha} \frac{\mu I_\beta}{2\pi r_{\alpha\beta}^2}\left(x_\alpha - x_\beta\right), \quad f_{\alpha y} = -I_\alpha \sum_{\beta \neq \alpha} \frac{\mu I_\beta}{2\pi r_{\alpha\beta}^2}\left(y_\alpha - y_\beta\right). \tag{4.12}$$

The force per unit length between two parallel current filaments is

$$\mathbf{f}_{12} = \frac{\mu I_1 I_2}{2\pi \left|\mathbf{r}_{12}\right|^2}\mathbf{r}_{12}. \tag{4.13}$$

where $\mathbf{f}_{12}$ is the force on filament 2, $\mathbf{r}_{12} = \left(x_1 - x_2, y_1 - y_2\right)$ is the position vector. Thus, we can define a potential energy between these two current filaments,

$$V_{12} = -\int_{r_{12}}^{\infty} \frac{\mu I_1 I_2}{2\pi r}\mathrm{d}r = -\frac{\mu I_1 I_2}{2\pi}\ln r_{12}, \tag{4.14}$$

and the total potential energy of the system is

$$V_e = -\frac{\mu}{4\pi}\sum_\alpha \sum_{\beta \neq \alpha} I_\alpha I_\beta \ln r_{\alpha\beta}. \tag{4.15}$$

The forces between parallel vortex filaments are derived in the Appendix. Comparing (4.12) with (A.9), we can find that $\left(-\mathbf{f}_\alpha\right)$ is analogous to $\mathbf{F}$, which keeps the same as the results in Section 2.4. For a two-dimensional point-vortex system, the following Hamiltonian can be defined [16],

$$h_f = -\frac{1}{4\pi} \sum_\alpha \sum_{\beta \neq \alpha} \Gamma_\alpha \Gamma_\beta \ln r_{\alpha\beta}, \tag{4.16}$$

which is similar to (4.15).

## 4.4 Magnetic moment and Vortical impulse

In electromagnetics, the magnetic moment is defined as[17]

$$\mathbf{m} = \frac{1}{2} \int (\mathbf{r} \times \mathbf{j}') \cdot dV, \tag{4.17}$$

where $\mathbf{j}'$ is the current density at the position $\mathbf{r}$. For a localized current distribution, the magnetic field induced by the a magnetic dipole with the magnetic moment $\mathbf{m}$ is

$$\mathbf{B}_d = -\frac{\mu}{4\pi} \nabla \left( \frac{\mathbf{m} \cdot \mathbf{r}}{|\mathbf{r}|^3} \right). \tag{4.18}$$

In hydrodynamics, the vortical impulse $\mathbf{I}_f$ is defined as[16]

$$\mathbf{I}_f = \frac{1}{2} \int (\mathbf{r} \times \boldsymbol{\omega}') \cdot dV. \tag{4.19}$$

where $\boldsymbol{\omega}'$ is the vorticity at $\mathbf{r}$. For the fluid flow without boundary, $\mathbf{I}_f$ stands for the total momentum of the fluid, which is a motion invariant. A localized vorticity density $\boldsymbol{\omega}$ gives rise to a velocity field as

$$\mathbf{V}_d = -\frac{1}{4\pi} \nabla \left( \frac{\mathbf{I}_f \cdot \mathbf{r}}{|\mathbf{r}|^3} \right), \tag{4.20}$$

where $\mathbf{I}_f$ can be seen as the strength of a dipole.

Therefore, we can see the correspondence between the magnetic moment and the vortical impulse.

## 4.5 Correspondence between electromagnetics and hydrodynamics

Here, we summarize the correspondence between electromagnetics and hydrodynamics in Table I.

Table I.    The correspondence between electromagnetics and hydrodynamics.

| Electromagnetics | Hydrodynamics |
|---|---|
| Magnetic induction intensity  $\mathbf{B}$ | Velocity  $\mathbf{v}$ |
| Current density $\mathbf{j}$ | Vorticity  $\boldsymbol{\omega}$ |
| Lorentz force density  $\mathbf{f} = \mathbf{j} \times \mathbf{B}$ | Lamb vector  $\mathbf{L} = \boldsymbol{\omega} \times \mathbf{v}$ |
| Current helicity | Hydrodynamic helicity |
| $H_e = \int \mathbf{j} \cdot \mathbf{B} \mathrm{d}V$ | $H_f = \int \boldsymbol{\omega} \cdot \mathbf{v} \mathrm{d}V.$ |
| Total current  $I = \dfrac{1}{\mu} \oint_C \mathbf{B} \cdot \mathrm{d}l$ | Circulation  $\Gamma = \oint_{C'} \mathbf{v} \cdot \mathrm{d}l$ |
| Magnetic energy density  $e_e = \dfrac{1}{2} \dfrac{\lvert \mathbf{B} \rvert^2}{\mu}$ | Kinetic energy density  $e_f = \dfrac{1}{2} \rho \lvert \mathbf{v} \rvert^2$ |
| Ampere force on a current filament | Lift force on an aerofoil |
| $\mathbf{F}_1 = -\mathbf{B}_1 \times \mathbf{i}$ | $\mathbf{F}_2 = \rho \mathbf{U}_1 \times \boldsymbol{\Gamma}$ |
| Magnetic moment | Vortical impulse |
| $\mathbf{m} = \dfrac{1}{2} \int (\mathbf{r} \times \mathbf{j}) \cdot \mathrm{d}V$ | $\mathbf{I}_f = \dfrac{1}{2} \int (\mathbf{r} \times \boldsymbol{\omega}) \cdot \mathrm{d}V$ |
| Electrical resistivity  $\chi$ | Fluid viscosity  $\eta$ |
| Current filament | Vortex filament |
| Current ring | Vortex ring |
| Superconductor | Solid |
| Maxwell equations+Ohm's law+Lorenz force | Navier-Stokes equation |

It should be noted that the electric field in the medium is negligible for the energy density and stress tensor. The only use of $\mathbf{E}$ is to drive the electric current. However, we show that $\mathbf{E}$ can be eliminated in the derivation of Navier-Stokes equation. Therefore, $\mathbf{E}$ is not crucial for the evolution of electromagnetic fields in the medium, and there is no need to find a physical variable corresponding to $\mathbf{E}$ in hydrodynamics. In a domain without boundary, the Biot-Savart formula of $\mathbf{E}$ can be derive from (2.18b) and (2.20) as

$$\mathbf{E} = \frac{1}{4\pi} \int \left( -\frac{\partial}{\partial t} \mathbf{B}' \right) \times \frac{\mathbf{r}}{|\mathbf{r}|^3} \mathrm{d}V, \tag{4.21}$$

where $\mathbf{B}'$ is the magnetic induction intensity at $\mathbf{r}$.

## 4.6 The basic concepts

The basic concept of hydrodynamics is the fluid particle, which is a mass point and can be traced in the motion. On the other hand, electromagnetics is based on the concept of field, which cannot be further explained by particles in mechanics. These two concepts lead to many differences. For example, the material derivative of a physical variable $\mathbf{T}$ in hydrodynamics is

$$\frac{\mathrm{D}\mathbf{T}}{\mathrm{D}t} = \frac{\partial \mathbf{T}}{\partial t} + (\mathbf{v} \cdot \nabla)\mathbf{T}. \tag{4.22}$$

However, there is no such definition in electromagnetics. The reason is that $\mathbf{T}$ is defined on a fluid particle, whose position is a function of time, so $\mathbf{T} = \mathbf{T}(\mathbf{x}(t), t)$. On the contrary, the physical variable of electromagnetic field has $\mathbf{A} = \mathbf{A}(\mathbf{x}, t)$, where $\mathbf{x}$ and $t$ are independent.

The basic concepts and the premises of governing equations are very different in electromagnetics and hydrodynamics. However, if we examine the fluid flow by using the Euler method, where the evolution of field variables are only concerned, then we can find many similarities of equations and patterns between these two subjects. It profoundly shows the unity of nature and also promote our understanding of physical reality.

## 5. Conclusion

In this paper, we investigate electromagnetics and hydrodynamics by making physical analogy between these two domains. The electromagnetic fields in a conducting medium are compared to the flow fields of an incompressible Newtonian fluid. The similarities in the governing equations and patterns between electromagnetic and flow fields are displayed.

It is found that the magnetic induction, current density, Lorenz force, superconductor boundary, Ohm's law and Ampere force in electromagnetics are analogous to the velocity, vorticity, Lamb vector, solid boundary, Newton's law of viscosity and Kutta-Joukowski theorem of lift force in hydrodynamics, respectively. We obtain the Navier-Stokes equation when the evolution of magnetic field in a conducting medium is examined by using the Maxwell equations, Lorenz force and Ohm's law. The result shows a profound relationship between electromagnetics and hydrodynamics, which can deepen our knowledge in these two fields.

## Appendix. The forces between parallel vortex filaments

When some parallel rotating filaments are imposed at fixed points in a fluid, the steady flow induced by these filaments can be described by the complex potential $w = \phi + \mathrm{i}\varphi$ in a complex plane $z = x + \mathrm{i}y$,

$$w(z) = \sum_{k=1}^{n} \frac{\varGamma_k}{2\pi \mathrm{i}} \ln(z - z_k), \tag{A.1}$$

where $\varGamma_k$ is the circulation of the point vortex at $z_k$. Here, the point vortex is an ideal model for the flow around a rotating filament and $z_k$ is the center of the $k$-th filament. The complex velocity $\upsilon = u - \mathrm{i}v = \mathrm{d}w/\mathrm{d}z$ of the point vortex at $z_j$ is

$$\frac{\mathrm{d}}{\mathrm{d}t} \overline{z}_j = \upsilon_{z_j} = \sum_{k \neq j} \frac{\varGamma_k}{2\pi \mathrm{i}} \frac{1}{z_j - z_k}. \tag{A.2}$$

The force per unit length on the rotating filament is

$$\mathbf{F} = F_x \mathbf{e}_x + F_y \mathbf{e}_y = -\oint_C p\mathbf{n}\mathrm{d}s, \tag{A.3}$$

where $\mathbf{e}_x, \mathbf{e}_y$ are the orthogonal unit vectors in the plane, C is a closed curve near the surface of filament and $\mathbf{n}$ is the outward unit normal vector. The complex force can be defined as follows,

$$F = F_x - \mathrm{i}F_y = -\mathrm{i}\oint_C p\mathrm{d}\overline{z}. \tag{A.4}$$

When the positions of rotating filaments are fixed, the flow induced by these filaments is steady and irrotational, so the Bernoulli theorem can be used and the complex force can be expressed by the Blasius formula[23],

$$F_x - \mathrm{i}F_y = \frac{\mathrm{i}\rho}{2} \oint_C \left(\frac{\mathrm{d}w}{\mathrm{d}z}\right)^2 \mathrm{d}z. \tag{A.5}$$

As

$$\left(\frac{\mathrm{d}w}{\mathrm{d}z}\right)^2 = \left(\sum_{k=1}^{n} \frac{\varGamma_k}{2\pi \mathrm{i}} \frac{1}{z - z_k}\right)^2, \tag{A.6}$$

there is only one singularity $z = z_j$ inside the loop C, we are only interested in the terms useful for the residue, so

$$\oint_C \left( \sum_{k=1}^{n} \frac{\Gamma_k}{2\pi i} \frac{1}{z - z_k} \right)^2 \mathrm{d}z = \oint_C \sum_{k \neq j} 2 \frac{\Gamma_j}{2\pi i} \frac{\Gamma_k}{2\pi i} \frac{1}{z - z_k} \frac{1}{z - z_j} \mathrm{d}z$$

$$= 2 \frac{\Gamma_j}{2\pi i} \sum_{k \neq j} \frac{\Gamma_k}{2\pi i} \frac{1}{z_j - z_k} \oint_C \left( \frac{1}{z - z_j} - \frac{1}{z - z_k} \right) \mathrm{d}z = 2 \Gamma_j \sum_{k \neq j} \frac{\Gamma_k}{2\pi i} \frac{1}{z_j - z_k}. \tag{A.7}$$

Thus, the complex force is

$$F_x - i F_y = \frac{i\rho}{2} \oint_C \left( \frac{\mathrm{d}w}{\mathrm{d}z} \right)^2 \mathrm{d}z = i\rho \Gamma_j \upsilon_{z_j}. \tag{A.8}$$

Substituting (A.2), we can derive that

$$F_x = \rho \Gamma_j \sum_{k \neq j} \frac{\Gamma_k}{2\pi r_{jk}^2} \left( x_j - x_k \right), \quad F_y = \rho \Gamma_j \sum_{k \neq j} \frac{\Gamma_k}{2\pi r_{jk}^2} \left( y_j - y_k \right). \tag{A.9}$$

**Acknowledgments**

This work has been supported by the National Natural Science Foundation of China (No.11872032), Zhejiang Provincial Natural Science Foundation (LY21A020006) and K. C. Wong Magna Fund in Ningbo University.

**Declaration of competing interest**

We declare that we have no financial and personal relationships with other people or organizations that can inappropriately influence our work, there is no professional or other personal interest of any nature or kind in any product, service and company that could be construed as influencing the position presented in or the review of the manuscript.

**References：**

University Press, 2000.